\documentclass[%
 reprint,
preprint,
nofootinbib,
amsmath, amssymb,
onecolumn,
]{revtex4-2}

\usepackage{bm,latexsym,mathrsfs,amssymb,amsmath}
\usepackage{graphicx}% Include figure files
\usepackage{dcolumn}% Align table columns on decimal point
\usepackage{bm}% bold math
\usepackage{inputenc}
\usepackage[breaklinks=true,unicode=true,urlcolor = blue,colorlinks = true,citecolor = blue,linkcolor = blue]{hyperref}
\usepackage{setspace} % To set line spacing
\usepackage{parskip}

\usepackage{geometry}
\begin{document}

\setstretch{2} 

\title{Microwave Resonances of Magnetic Skyrmions in\\  Thin Film Multilayers}% Force line breaks with \\
%\thanks{A footnote to the article title}%

\author{Bhartendu Satywali$^{1,\star}$, Volodymyr P. Kravchuk$^{2,3,\star}$, Liqing Pan$^4$, M. Raju$^1$,  Shikun He$^1$, Fusheng Ma$^{1}$,  \\ A.P. Petrovi{\'c}$^{1,\dagger}$,  Markus Garst$^{3,5,6}$, Christos Panagopoulos$^{1,\dagger}$}

\affiliation{$^1$Division of Physics and Applied Physics, School of Physical and Mathematical Sciences,\\ Nanyang Technological University, 637371 Singapore.\\
$^2$Bogolyubov Institute for Theoretical Physics of National Academy of Sciences of Ukraine, 03680 Kyiv, Ukraine.\\
$^3$Institute for Theoretical Solid State Physics, Karlsruhe Institute of Technology, 76131 Karlsruhe, Germany.\\
$^4$Research Institute for Magnetoelectronics and Weak Magnetic Field Detection, College of Science, China Three Gorges University, 443002 Yichang, China.\\
$^5$Institut f\"ur Theoretische Physik, TU Dresden, 01062 Dresden, Germany.\\
%$^6$School of Physics and Technology, Nanjing Normal University, 210046 Nanjing, China.\\
$^6$Institute for Quantum Materials and Technologies, Karlsruhe Institute of Technology, 76021 Karlsruhe, Germany.\\
$\star$ Equal Contribution, $\dagger$ Corresponding Authors \\
appetrovic@ntu.edu.sg, christos@ntu.edu.sg
}

%\date{\today}% It is always \today, today,
             %  but any date may be explicitly specified

\maketitle

\setstretch{1.5} 

%\textcolor{red}{Labeling of Figure 5 needs to be adjusted to comply with Referee 3.}
%\textcolor{red}{Linecuts in Figure 3e and f need to be adjusted to comply with Referee 3.}

\section*{Abstract}
\textbf{Non-collinear magnets exhibit a rich array of dynamic properties at microwave frequencies. They can host nanometer-scale topological textures known as skyrmions, whose spin resonances are expected to be highly sensitive to their local magnetic environment. Here we report a magnetic resonance study of an [Ir/Fe/Co/Pt] multilayer hosting N\'eel skyrmions at room temperature. Experiments reveal two distinct resonances of the skyrmion phase during in-plane ac excitation, with frequencies between 6-12 GHz. Complementary micromagnetic simulations indicate that the net magnetic dipole moment rotates  counterclockwise (CCW) during both resonances.  The magnon probability distribution for the lower-frequency resonance is localised within isolated skyrmions, unlike the higher-frequency mode which principally originates from areas between skyrmions. However, the properties of both modes depend sensitively on the out-of-plane dipolar coupling, which is controlled via the ferromagnetic layer spacing in our heterostructures. The gyrations of stable isolated skyrmions reported in this room temperature study encourage the development of new material platforms and applications based on skyrmion resonances. Moreover, our material architecture enables the resonance spectra to be tuned, thus extending the functionality of such applications over a broadband  frequency range.}

\section*{Introduction}
Magnetic skyrmions are stabilised due to competition between various spin interactions including Heisenberg exchange, the Dzyaloshinskii-Moriya interaction (DMI) and geometric frustration. These non-collinear configurations of magnetic moments have been envisioned as a platform for the next generation of spintronic devices. An essential requirement to realise this technological ambition is a complete understanding of the static and dynamic properties of skyrmions.\\
The majority of existing literature on skyrmion dynamics is restricted to non-centro\-symmetric single crystals hosting either Bloch or Néel skyrmion lattices at low temperatures. These are magnetic materials with space group P2$_1$3 (e.g. Cu$_2$OSeO$_3$, MnSi or Fe$_{1-x}$Co$_x$Si), or R3m (e.g. GaV$_4$S$_8$) \cite{Seki2016a,Garst2017}. In both types of system, the skyrmions exhibit clockwise (CW) and counterclockwise (CCW) gyration modes as well as a breathing (BR) mode \cite{Mochizuki2012,Onose2012,Okamura2013,Schwarze2015,Ehlers2016, Zhang2016}. 
In contrast with magnetic skyrmion arrays in bulk single crystals, 
Néel skyrmions at ferromagnet-heavy metal (FM-HM) interfaces are stabilized by the interfacial Dzyaloshinskii-Moriya interaction. Their configuration can range from dilute to dense disordered arrays, as well as ordered lattices. However, the skyrmion size, stability and configurations are all sensitive to the magnetic field history and  uniformity of the interface. Furthermore, additional energy contributions (such as long-range magnetostatic interaction between separate ferromagnetic layers) can also play a decisive role in determining the skyrmion properties, by modifying the  helicity \cite{Legrand2018}. Magnetic multilayers are particularly promising candidates for developing skyrmion-based technologies, since they combine skyrmion stability at room temperature with the ability to finely tune their magnetic parameters via the multilayer geometry. The latter significantly amplifies the role of the dipole-dipole interaction, enriching the properties of skyrmion phases.  \\
Recent experiments on ferromagnet-heavy metal (FM-HM) multilayers have yielded considerable insight into the static properties of N\'eel skyrmions~\cite{Soumyanarayanan2017, Romming2013, Moreau2016, Zeissler2018, Raju2019} and their motion due to dc electrical currents~\cite{Woo2016, Litzius2017, Jiang2017} or magnetic field gradients~\cite{Zhang2018a}. However, the dynamic properties of Néel skyrmions in FM-HM heterostructures have so far resisted investigation, due to the reduced sensitivity of magnetic resonance probes in thin films and the difficulty of engineering high skyrmion densities. Consequently, many interesting and potentially useful predictions for skyrmion resonances~\cite{Mochizuki2012, Mochizuki2013, Lin2014b, Schutte2014a, Wang2015, Kravchuk2018}, such as RF energy harvesters, microwave couplers and magnon gratings~\cite{ Finocchio2015,  Finocchio2016, Garst2017}, remain to be experimentally verified. A clear understanding of skyrmion responses at GHz frequencies is also prerequisite for operating recently-postulated synaptic computational architectures at useful clock frequencies~\cite{Li2017, Prychynenko2018, Song2020}. Detailed experimental and numerical studies of skyrmion dynamics will be essential to deliver the many proposed applications requiring skyrmion interactions with radio or microwave frequency fields.
\\
Here we present a broadband microwave absorption study of sputter-deposited multilayers hosting stable N\'eel skyrmions at room temperature and in out-of-plane (OP) fields 140 mT~$ \lesssim \mu_0H_\perp  \lesssim $~300 mT. Applying additional in-plane (IP) microwave fields in the range 5-15 GHz, we detect three resonant modes which correlate with the evolving magnetisation configurations in our samples. Our micromagnetic simulations indicate that the low frequency (LF) mode is associated with a CCW gyration of the skyrmion core, while the CCW precession of the magnetisation within the inter-skyrmion zones is the predominant contributor to the high frequency (HF) mode. The intensity and resonant frequencies of both HF and LF modes are acutely sensitive to the interlayer dipolar interaction between ferromagnetic layers, which can be modulated by the thickness of the non-magnetic spacer layer. The third mode emerges at high fields and is associated with the uniform precession of the field-polarised magnetisation.\\
\begin{figure*}[t!]
\includegraphics[width=\textwidth]{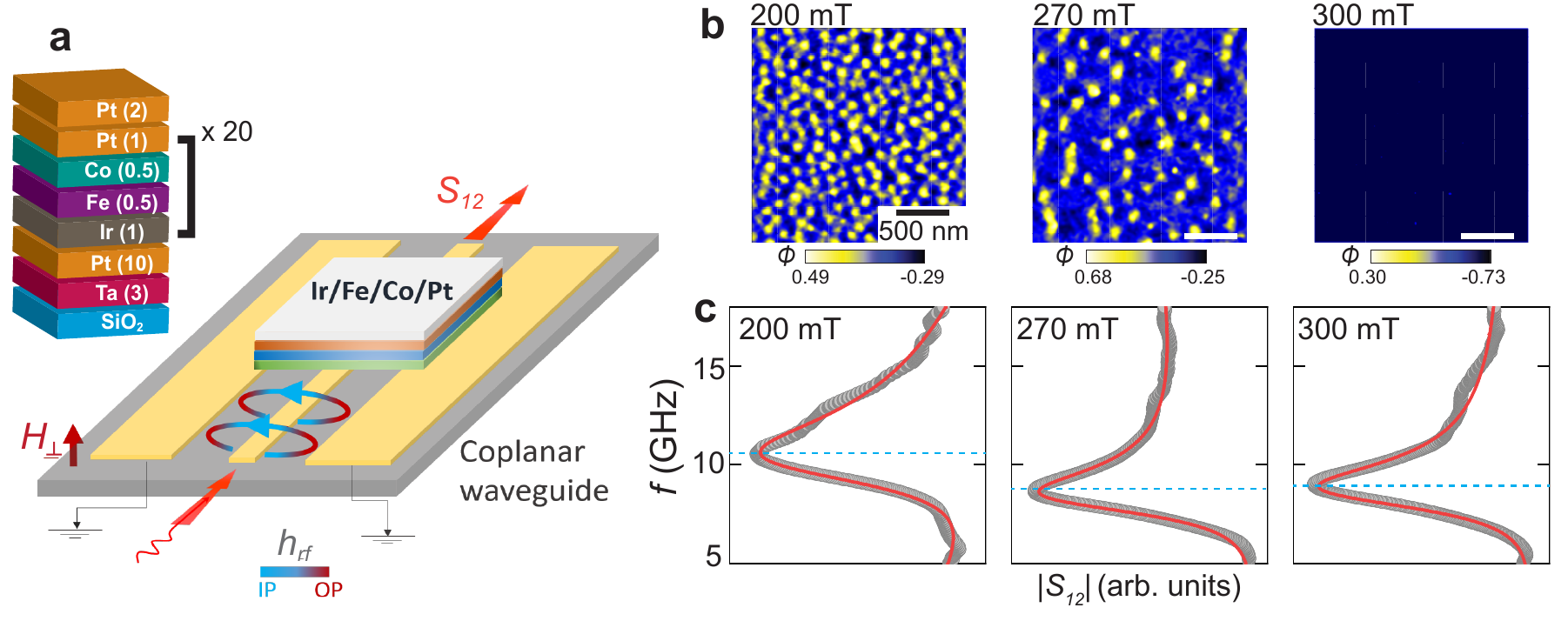}
\caption{\textbf{Characterising resonances across the magnetic phase diagram of  [Ir\textsubscript{1}Fe\textsubscript{0.5}Co\textsubscript{0.5}Pt\textsubscript{1}]\textsuperscript{20}.} \textbf{a} Experimental geometry: a microwave signal applied to the coplanar waveguide sample-holder creates an oscillating in-plane (IP) magnetic field $h_{rf}$ in our [Ir\textsubscript{1}Fe\textsubscript{0.5}Co\textsubscript{0.5}Pt\textsubscript{1}]\textsuperscript{20} heterostructure, which lies perpendicular to the out-of-plane (OP) dc field $H_\perp$. The absolute value of the microwave transmission through the waveguide, $S_{12}$, is measured using a vector network analyser in two-port mode. Magnetic resonance within the sample leads to an increased microwave absorption and hence a reduction in $S_{12}$. The inset shows a cross-sectional view of the multilayer stack. \textbf{b} Magnetic force microscopy (MFM) images of  [Ir\textsubscript{1}Fe\textsubscript{0.5}Co\textsubscript{0.5}Pt\textsubscript{1}]\textsuperscript{20} acquired at temperature 300\,K in dc fields of 200\,mT, 270\,mT and 300\,mT. At these fields, the sample exhibits a dense skyrmion array, isolated skyrmions and ferromagnetic order, respectively.  The colour bar shows the phase shift ($\phi$, in degrees) of the MFM cantilever oscillations, which is proportional to the OP magnetisation component $m_z$. \textbf{c} Frequency-sweep microwave absorption spectra acquired at the same three magnetic fields. Resonance is visible as a local minimum in $S_{12}$  at each field (blue dashed lines). We extract the resonant frequency dispersions $f( H_\perp)$ using a standard Dysonian peak-fitting routine~\cite{Dyson1955} (red lines) on raw $ S_{12}(f, H_\perp)$ absorption spectra (grey lines).}
\label{fig:1}
\end{figure*}
\section*{Results}
\begin{figure*}[t!]
\includegraphics[width=\textwidth]{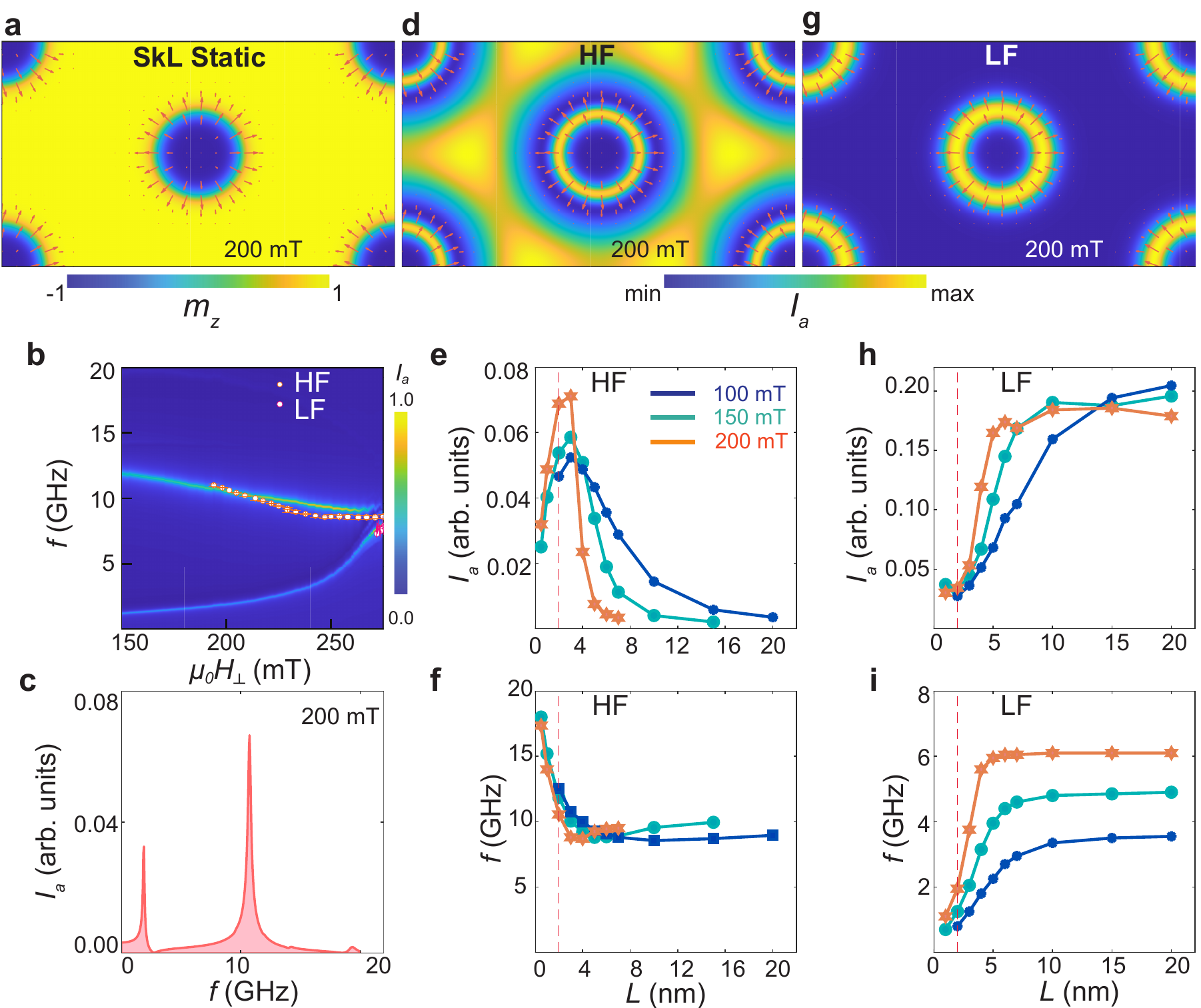}
\caption{\textbf{Skyrmion lattice simulations exposing the delocalised HF resonance.} We use a rectangular unit cell in a hexagonal skyrmion lattice containing two skyrmions, with area $S_{uc}=\sqrt{3} a^2$. The lattice constant $a$ is adjusted such that the topological charge density $\rho_{top}=\frac{2}{S_{uc}}$ coincides with that of a large (2$\times$2\,$\mu$m\textsuperscript{2}) relaxed sample at the same field (Supplementary section II.C, Fig. S9). \textbf{a} Equilibrium magnetisation at 200\,mT for a unit cell size 222$\times$128\,nm\textsuperscript{2}. The colour scale and arrows represent the out-of-plane and in-plane magnetisations.   \textbf{b} Simulated resonance spectra for the geometry described in \textbf{a}.  \textbf{c} displays a vertical linecut through the spectra in \textbf{b} at 200 mT, highlighting the relative intensity of the HF and LF modes. \textbf{d}  and \textbf{g} display the spatial distribution of the microwave absorption intensity during HF and LF resonances at 200\,mT, excited at 10.6\,GHz and 2\,GHz, respectively. While the distribution of the LF mode is localised close to the skyrmion core, the HF mode intensity is primarily concentrated in the inter-skyrmion area.  Panels \textbf{a} – \textbf{d} and \textbf{g} are simulated with a  spacer thickness $L$ = 2\,nm. Panels \textbf{e}, \textbf{f} and \textbf{h}, \textbf{i} depict the variation of the absorption intensities and resonant frequencies with $L$ for the HF and LF modes, respectively. The red dashed line at $L$ = 2\,nm corresponds to the experimental layer separation.}
\label{fig:5}
\end{figure*}

Our samples are [Ir\textsubscript{1}Fe\textsubscript{0.5}Co\textsubscript{0.5}Pt\textsubscript{1}]\textsuperscript{20} multilayers grown by magnetron sputtering, where the subscripts are the sequential layer thicknesses in nanometers, and the complete “stack” comprises 20 repeats of [Ir\textsubscript{1}Fe\textsubscript{0.5}Co\textsubscript{0.5}Pt\textsubscript{1}]. This stack is known to host N\'eel skyrmions from room temperature down to 5\,K~\cite{Raju2019, Yagil2018}. To probe the electrodynamic properties of these stacks, we employ a broadband technique~\cite{He2016, He2017, Okada2017} and measure the microwave transmission $S_{12}$ through a proximate coplanar waveguide (Fig.~\ref{fig:1}a). 
\begin{figure*}[t!]
\includegraphics[width=\textwidth]{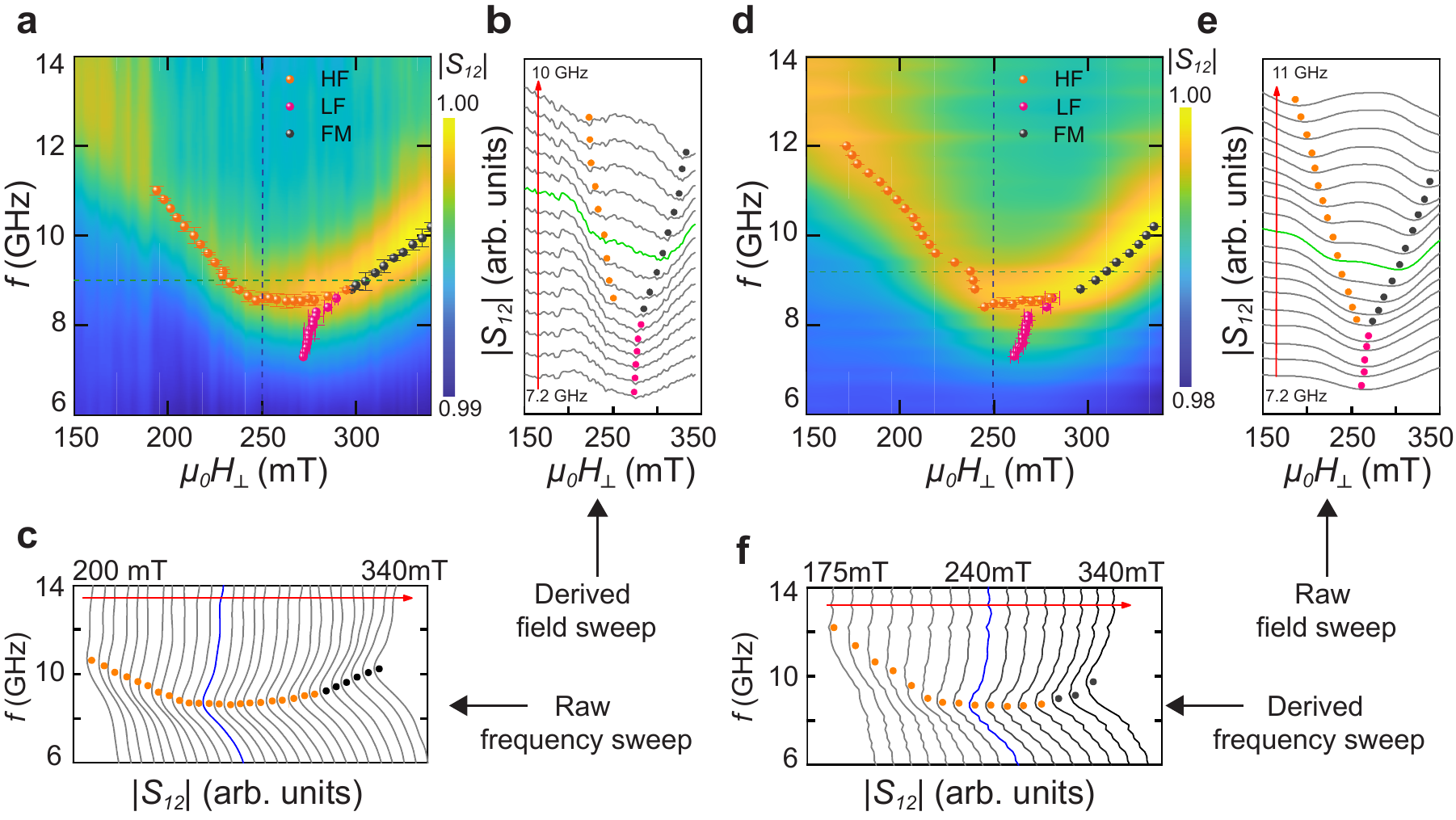}
\caption{\textbf{Experimental microwave absorption spectra of  [Ir\textsubscript{1}Fe\textsubscript{0.5}Co\textsubscript{0.5}Pt\textsubscript{1}]\textsuperscript{20} measured by frequency and field sweep  ferromagnetic resonance techniques at room temperature.} \textbf{a} Transmission intensity $S_{12} (f, H_\perp)$ (colour map) determined from frequency-swept $S_{12} (f)$ curves at a series of constant magnetic fields. Resonances (overlaid data-points) are determined by Dysonian peak-fits in raw  $S_{12} (f)$  curves (\textbf{c}), or Lorentzian peak-fits in interpolated  $S_{12} (H_\perp)$ curves (\textbf{b}). Black data-points track resonances in the ferromagnetic (FM) phase. In the skyrmion phase, the high frequency (HF) and low frequency (LF) modes are shown in orange and pink respectively. Blue  $S_{12} (f)$ and green  $S_{12} (H_\perp)$ curves are cuts along the blue and green dashed lines in the intensity plot. \textbf{d} Similar plots showing  $S_{12} (f, H_\perp)$  obtained from a separate dataset consisting of field-swept  $S_{12} (H_\perp)$ curves at a series of constant frequencies. Raw $S_{12}( H_\perp)$ curves are shown in \textbf{e}, and derived frequency-swept $S_{12}(f)$ curves at constant field in \textbf{f}. The error bars in a, d are obtained from least-square fits of raw experimental spectra. Details of the fitting procedure are included in supplementary section I.B.}
\label{fig:2}
\end{figure*}
Adjusting the OP dc field $\mu_0H_\perp$ tunes our multilayers through a sequence of magnetic orders, revealed by magnetic force microscopy (MFM) imaging (Fig.~\ref{fig:1}b).  For $\mu_0H_\perp \gtrsim$ 300\,mT the magnetisation is saturated, but a slight decrease in field below $\sim$300\,mT results in skyrmion nucleation at random sites.  As we continue to reduce the field, further skyrmions are progressively nucleated, transforming the magnetic configuration into a densely-packed disordered skyrmion lattice which persists down to $\sim$140\,mT. 
The disordered configuration most likely originates from local inhomogeneities in the magnetic interactions, due to structural defects and a complex energy landscape. Microwave resonances can be identified in all applied fields as local minima in the frequency-dependent transmission $S_{12} (f)$ (Fig.~\ref{fig:1}c).  \\
Figure~\ref{fig:2}a depicts the microwave absorption across the phase diagram, measured via frequency-sweep spectroscopy at constant field. Searching for local minima in cuts along both field (Fig.~\ref{fig:2}b) and frequency  (Fig.~\ref{fig:2}c) axes we identify three distinct resonant branches which correlate with the different magnetic configurations imaged via MFM. In the high-field polarised ferromagnetic  state, the Kittel resonance displays the expected linear $f(H_\perp)$ relation. Following skyrmion nucleation below $\sim$300\,mT, the Kittel mode splits into two branches:  a narrow resonance which softens rapidly, and a broad absorption which retains an approximately constant frequency as the field is reduced to $\sim$250\,mT, below which the resonance frequency rises sharply with decreasing field. We label these branches LF and HF, respectively. For comparison, Fig.~\ref{fig:2}d shows absorption spectra from a similar heterostructure during field-sweep measurements. Despite the known sensitivity of skyrmion configurations to the magnetic field history~\cite{Duong2019}, the results are remarkably consistent, with all three resonances emerging at similar frequencies in both experiments.  The LF mode cannot be resolved in linecuts along the frequency axis in either dataset (Fig.~\ref{fig:2}c,f), due to its relatively low absorption intensity and steep frequency dispersion with magnetic field.  However, it is clearly visible in field axis linecuts (Fig.~\ref{fig:2}b,e) for data acquired in frequency and field-swept measurements.  Our results therefore indicate the robust nature of both LF and HF resonances in the skyrmion phase.\\
\begin{figure*}[t!]
\includegraphics[width=\textwidth]{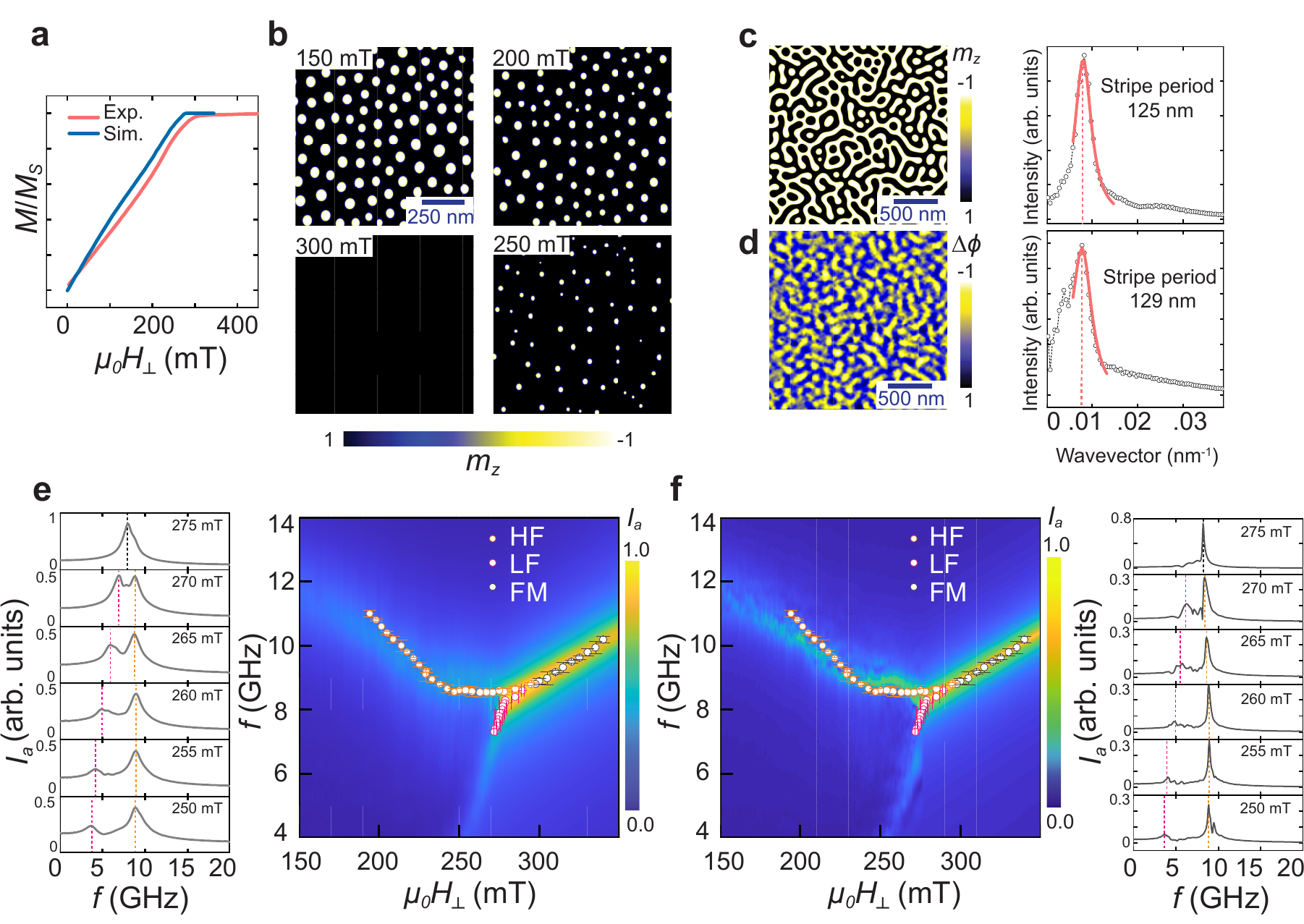}
\caption{\textbf{Micromagnetic simulations of a [Ir\textsubscript{1}Fe\textsubscript{0.5}Co\textsubscript{0.5}Pt\textsubscript{1}]\textsuperscript{20}multilayer.} The simulation model consists of a ferromagnetic-nonmagnetic [FM\,(1\,nm) - NM\,(2\,nm)] bilayer of size 2$\times$2\,$\mu$m\textsuperscript{2} constructed by grids of size 3.9$\times$3.9$\times$1\,nm\textsuperscript{3}, with periodic boundary conditions (PBCs) applied along the $x$, $y$ and $z$ directions. We use the following experimentally determined magnetic parameters: exchange stiffness $A$ = 9.25\,pJ/m, Dzyaloshinskii-Moriya interaction $D$ = 1.40\,mJ/m\textsuperscript{2}, uniaxial anisotropy $K$ = 0.65\,MJ/m\textsuperscript{3} and saturation magnetisation $ M_s $ = 1.02\,MA/m. \textbf{a} Simulated OP magnetisation as a function of OP dc field $H_\perp$, compared with our experimental results. \textbf{b} Simulated spatial OP magnetisation maps for increasing OP dc field. \textbf{c} Simulated labyrinthine stripe phase at zero field. \textbf{d} Zero field stripe phase experimentally imaged by magnetic force microscopy (MFM). The colour scale represents the normalised phase shift $\Delta\phi$ in the MFM cantilever oscillations. Discrete Fourier transforms  of the simulated and experimental magnetisations yield similar stripe periodicities. \textbf{e} Microwave absorption intensity $I_a(f, H_\perp)$ simulated with the experimentally 
 estimated damping {\bf $\alpha = 0.05$}. The spectra are calculated by applying a sinc pulse of an excitation field $\textbf{b}(t)$ = $\hat{\textbf{e}}_x b_0$sinc$( 2\pi f_{max} (t-t_0)$) centered at $t_0 =$ 1\,ns with  $f_{max}$ = 20\,GHz and $b_0$ = 10\,mT, then Fourier transforming the induced IP magnetisation oscillations $m_x$. The overlaid data points are the resonances from frequency sweep experiments at 300\,K. Resonance spectra obtained from vertical (frequency-swept) line-cuts are also shown, where dashed lines highlight the positions of the local maxima.  \textbf{f} Absorption intensity spectra and vertical line-cuts simulated with  a lower damping, $\alpha$ = 0.01, to clearly expose the individual resonances. The dashed lines marking the positions of the LF (pink) and HF (orange) peaks in the linecut graphs in \textbf{e}, \textbf{f} are visual guides highlighting absorption maxima.}
\label{fig:3}
\end{figure*}
To understand the microscopic origin of the experimentally observed LF and HF modes, we perform micromagnetic simulations using MuMax\textsuperscript{3}~\cite{Vansteenkiste2014}. The magnetic configuration at each field was obtained by relaxing from a random magnetisation. Close to zero field, the magnetisation relaxes into a labyrinth stripe structure, which transforms into a dense skyrmion phase for $ \mu_0H_\perp  \approx$ 140\,mT. The evolution of the simulated OP magnetisation $m_z$ lies within 30\,mT of our experimental results, as shown in Fig.~\ref{fig:3}a, confirming the accuracy of our simulation parameters. Slight deviations between the two datasets are to be expected since thermal fluctuations and spatial inhomogeneities (which may influence relaxation processes) are not considered in the present simulations. The spatial magnetic configuration is depicted in Fig.~\ref{fig:3}b. At intermediate fields even this clean system relaxes into a disordered skyrmion lattice, emphasising the complex energy landscape which can easily trap skyrmions in metastable configurations. With increasing field, the skyrmion array gradually transforms from a disordered lattice into a dilute gas of isolated skyrmions and, eventually, into a field-polarised magnetic state, in accordance with our MFM imaging (Fig.~\ref{fig:1}b). The validity of our simulation parameters - especially the exchange $A$ and DMI $D$ - is further confirmed by the close agreement of the simulated zero-field stripe periodicity $\sim$ 125\,nm with the experimentally measured value $\sim$129\,nm (Fig.~\ref{fig:3}c,d). 
\\
 The simulated microwave absorption spectrum as a function of field and frequency is shown in Fig.~\ref{fig:3}e for an intrinsic damping parameter $\alpha = { 0.05}$ determined from fits to our experimental data (Supplementary section I.A). Three distinct resonance modes can be identified in the simulated spectra, in agreement with experiments. The frequency sweeps indicate that the broad linewidth leads to a superposition of resonance signals, thus hampering a clear delineation of the LF and FM modes. For this reason, we performed further simulations with a lower damping parameter $\alpha$ = 0.01 (Fig.~\ref{fig:3}f), allowing us to resolve the dispersion of all the modes. We note that the simulated absorption intensity of the LF mode is higher than in  our experimental data (Fig.~\ref{fig:2}).  This can be attributed to the presence of extrinsic, mode-dependent damping mechanisms in our heterostructures which are not accounted for in our simulations (see Supplementary section I.B for further discussion).  Nevertheless, the field-dependent resonant frequencies of all three modes are in close agreement with our experimental findings.  As discussed below in more detail, the interlayer dipolar coupling is essential to reproduce our experimental results, emphasising the critical role of dipolar interactions in multilayers.\\
\begin{figure*}[t!]
\includegraphics[width=\textwidth]{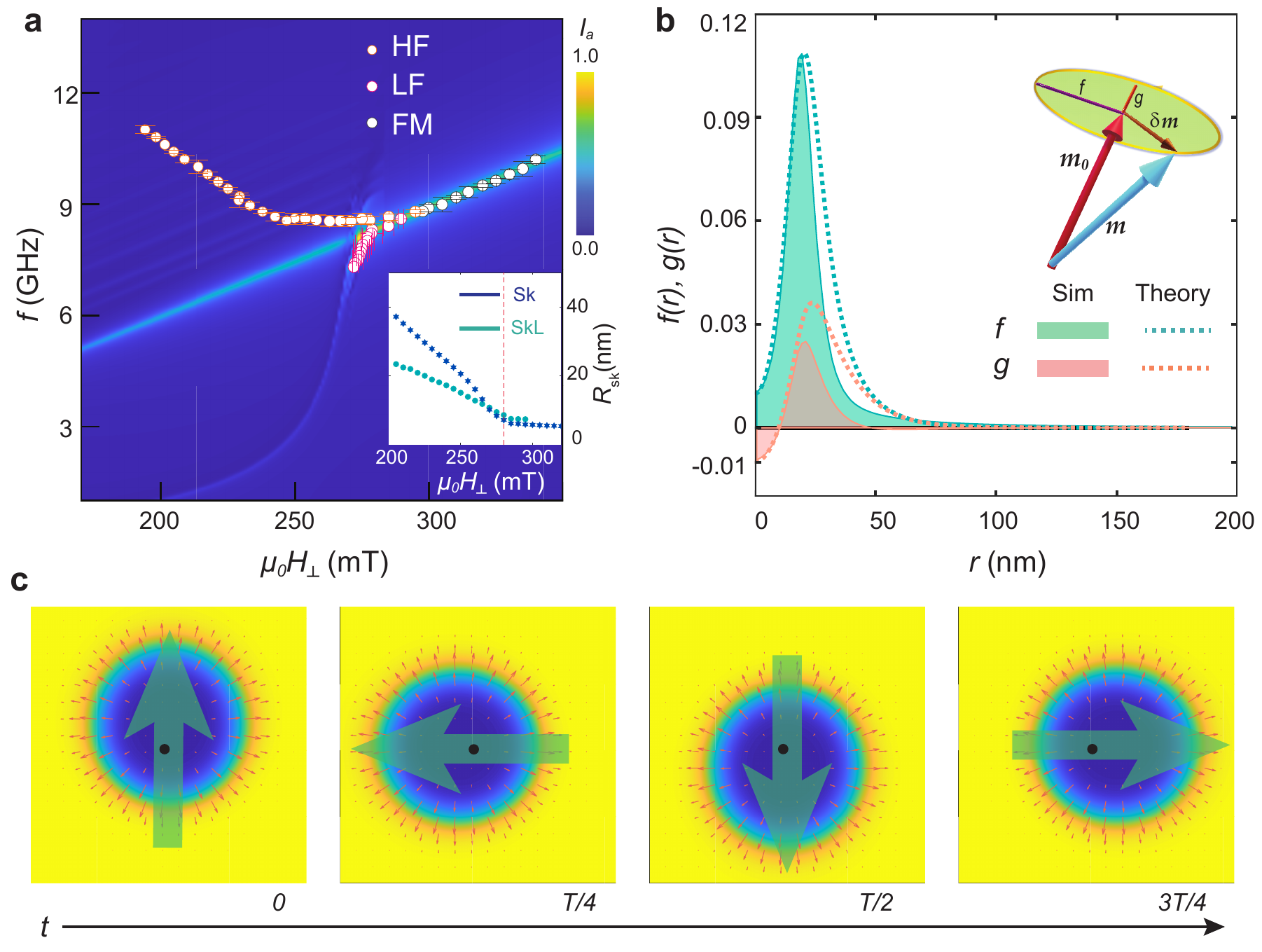}
\caption{\textbf{Single-skyrmion simulations exposing the counter-clockwise (CCW) nature of the LF resonance}. Micromagnetic simulations were performed on a prepared metastable state of area 1$\times$1\,$\mu$m\textsuperscript{2}  consisting of a single skyrmion in a field-polarised background.  \textbf{a} Simulated absorption intensity obtained from the Fourier transform of the in-plane magnetisation $ m_x (t) $ after excitation by an in-plane sinc pulse with amplitude $b_0$ = 2\,mT.  The overlaid symbols show our experimental data for comparison. The inset depicts the field-dependent radius of the isolated skyrmion (Sk) compared with the skyrmion radius in the regular lattice (SkL), discussed in Fig.~\ref{fig:5}. The red dashed line at  $ \mu_0H_\perp$= 280\,mT marks the onset of the LF resonance identified as a CCW magnon-skyrmion bound state.  \textbf{b} Spin wave eigenfunctions $f(r)$, $g(r)$  of the CCW resonance at $ \mu_0H_\perp$= 250\,mT,  respectively parametrised by the semi-major and semi-minor axes of the precessional ellipse of the local magnetisation. Our numerical data (Sim) is compared to an analytical approximation (Theory). \textbf{c} Temporal evolution of the CCW skyrmion resonance at  $ \mu_0H_\perp $= 250\,mT with oscillation period T = 345\,ps, where the large green arrow represents the total dipole moment.  The black dot represents the stationary centre defined by the first moment of the topological charge density $\rho_{top}=\frac{1}{4\pi} \textbf{m} \cdot(\partial_x \textbf{m} \times \partial_y \textbf{m})$. The size of the displayed area is 78$\times$78\,nm\textsuperscript{2}.}
\label{fig:4}
\end{figure*}
To elucidate the character of the skyrmionic excitation modes we performed further simulations on two additional configurations. In the first instance, a metastable state was prepared consisting of a single skyrmion in a field-polarised background. The absorption intensity of this specific state and its evolution with $ \mu_0H_\perp$ is displayed in Fig.~\ref{fig:4}a. For the full range of applied fields 150\,mT $\leq \mu_0 H_\perp \leq $ 350\,mT, the field-polarised background contributes to the spectral weight of the Kittel resonance, corresponding to the uniform precession of the field-polarised magnetic moments. At $ \mu_0H_\perp =$ 280\,mT, the size of the skyrmion suddenly increases with decreasing field, coinciding with the emergence of an additional resonant mode below the Kittel frequency. This resonance corresponds to a magnon-skyrmion bound state with CCW character (Fig.~\ref{fig:4}c) whose frequency tracks the LF resonance observed experimentally. Its spin wave eigenfunctions extracted from the numerical data can be parametrised by the semi-major and semi-minor axes of the local precessional ellipse, $f(r)$ and $g(r)$, respectively, and  agree well with analytical results (see Fig.~\ref{fig:4}b and Supplementary section III.C for details). Note that the sign of the product $f g$ determines the sense of rotation of the magnetisation vector $\bf {m}$ around its equilibrium direction. Importantly, the spin wave functions exhibit a maximum close to the skyrmion radius but then decay rapidly to zero at large distances, indicating the localisation of the LF resonance within the skyrmion area. The corresponding time evolution is illustrated in Fig.~\ref{fig:4}c, where the large arrow represents the total dipolar moment performing CCW gyration.  \\
Having identified the LF resonance as a CCW magnon-skyrmion bound state, we now discuss the HF resonance. The frequency of the HF mode is located above the Kittel resonance frequency and thus within the scattering continuum of the field-polarised background, which suggests that it cannot be accounted for by the first setup containing only a single skyrmion. The HF resonance is therefore likely to be associated with a multi-skyrmion configuration. We have verified that both HF and LF resonances observed in the simulations in Fig.~\ref{fig:3} persist after annealing the metastable disordered skyrmion configuration into a regularly ordered hexagonal lattice. The annealing was achieved by shaking the system with the help of a large amplitude ac magnetic field, using a similar approach  to vortex matter in superconductors \cite{Avraham2001} (see Supplementary section II.D). \\
In order to study the HF mode, we consider a second simulation configuration, consisting of a rectangular unit cell in a regular hexagonal lattice containing two skyrmions. Figure~\ref{fig:5}a shows the corresponding equilibrium magnetisation $\textbf{m}_0 (x,y)$ at 200 mT. The numerically-obtained absorption spectrum of this system is depicted in Fig.~\ref{fig:5}b, together with the experimentally observed frequencies. In the field range encompassed by the skyrmion phase, two distinct resonances can be identified whose field dependences closely resemble those of the HF and LF modes observed experimentally. Moreover, the spectral intensity of the HF mode is stronger than that of the LF mode (Fig.~\ref{fig:5}c), in agreement with experiment. The magnon probability distribution of the two modes is obtained numerically by time-averaging $\langle |\textbf{m} (x,y,t) - \textbf{m}_o (x,y)|^2 \rangle_t $ and shown in Fig.~\ref{fig:5}d,g. The distribution of the LF mode is localised and concentrated on a ring around each skyrmion, in agreement with the result of the single-skyrmion calculations of Fig.~\ref{fig:4}b. In contrast, the probability density of the HF mode is mainly distributed within the area between the skyrmion positions. The precessional motion of the polarised magnetic moments in the inter-skyrmion zones also results in a CCW rotation of the total magnetic dipole moment for the HF mode (illustrated in our Supplementary videos). 
\\
A characteristic advantage of multilayers compared to non-centrosymmetric bulk crystals and thin film monolayers is the tunability of the interlayer dipolar interaction by varying the stack geometry. To study the influence of interlayer dipolar interactions numerically, we varied the thickness $L$ of the nonmagnetic spacer layer, corresponding to the [Ir-Pt] thickness in our multilayers. Thicker layers result in a relatively weaker interlayer dipolar interaction. The evolution of the resonance frequencies  and absorption intensities as a function of $L$ are shown in Fig.~\ref{fig:5}e,f,h,i. We find that the resonances are highly sensitive to the dipolar coupling when $L$ is comparable to the thickness of the magnetic layer (1\,nm), but saturate at a thickness $L \geq$ 10\,nm, which is much smaller than the skyrmion diameter ($\sim$ 60 - 80\,nm) or the inter-skyrmion distance ($\sim$ 130 - 150\,nm). The resonant frequencies of the HF and LF modes respectively decrease and increase with $L$,  before saturating at larger thickness. The spectral weight of the HF mode possesses a characteristic maximum at $L$  $\sim$ 2-5\,nm depending on the applied field, before decreasing and eventually vanishing for large $L$. The HF mode therefore cannot be observed in the limit of large $L$ where the layers are practically decoupled. In contrast, the weight of the LF mode increases with $L$ and saturates at a finite intensity for large thickness. Whereas the spectral intensity of both modes increases with applied field for thinner spacers, this trend reverses for larger $L$. For our experimental thickness $L$ = 2\,nm the intensity of the HF mode is larger than that of the LF mode for all fields, in agreement with our experimental results. This evident sensitivity of skyrmion resonances to the geometry of chiral magnetic multilayers highlights their potential for use in tunable microwave receivers. 
\\

\section*{Discussion}
It is instructive to compare our results with previous work on bulk magnets with a Dzyaloshinskii-Moriya interaction~\cite{Mochizuki2012,Onose2012,Okamura2013,Schwarze2015,Ehlers2016, Zhang2016}. The magnetic skyrmion lattice of these systems at low temperatures is known to possess a CCW mode and a clockwise (CW) mode for IP ac magnetic fields, as well as a breathing mode for OP ac fields. 
Such behaviour is distinct from the spectra we measure for skyrmions in multilayers: in particular, there is  no CW mode within our experimental frequency range. This difference can be attributed to the presence of an uniaxial anisotropy $K$, which is absent in cubic chiral magnets, and - more importantly -
to the strong influence of the dipolar interaction. The relative strength of the latter is quantified by the dimensionless parameter $\mu_0 A M_s^2/D^2$, where $\mu_0$ is the vacuum permeability, $A$ the exchange stiffness, $D$ the DMI and $M_s$ the saturation magnetisation. In our multilayers this parameter is one order of magnitude larger than in bulk chiral magnets hosting Bloch skyrmions \cite{Garst2017}. Moreover, N\'eel skyrmions possess finite volume magnetostatic charges, unlike Bloch skyrmions. These aspects drastically modify the excitation spectrum of skyrmion resonances for our magnetic multilayers in comparison with bulk chiral magnets. As a result, the CW mode has a significantly lower intensity and is shifted to high frequencies beyond our experimentally accessible frequency range. At the same time, the CCW excitation softens and is shifted to lower frequency due to this enhanced attraction of spin waves in the CCW channel. This produces the magnon-skyrmion bound state for a single skyrmion shown in Fig.~\ref{fig:4}, which is responsible for the LF resonance observed in our multilayers. 
A bound CCW state of this nature does not exist in cubic chiral magnets \cite{Schutte2014a}. In addition, due to the enhanced attraction in the CCW channel the next higher-order CCW mode (possessing an additional node in its wavefunction, as shown in Fig.~\ref{fig:5}d) becomes experimentally accessible, resulting in our observed HF mode. 
This pronounced sensitivity to dipolar interactions enables a remarkable tunability of the CCW resonances through the multilayer geometry.
%
%\textcolor{blue}{We attribute this difference to the strong influence of the dipolar interaction, whose relative strength is quantified by the dimensionless parameter $\mu_0 A M_s^2/D^2$ with the vacuum permeability $\mu_0$, the exchange stiffness $A$, the DMI $D$ and the saturation magnetization $M_s$. In our multilayers this parameter is one order of magnitude larger than in bulk chiral magnets hosting Bloch skyrmions \cite{Garst2017}. In addition, N\'eel skyrmion, in contrast to Bloch skyrmions, possess finite volume magnetostatic charges. These two aspects drastically modify the excitation spectrum of skyrmion resonances in our magnetic multilayer system as compared to the bulk chiral magnets. As a result, the CW mode has significantly lower intensities and is shifted to high-frequencies beyond the experimentally inaccessible frequency range (not shown). Instead, two CCW modes materialize in our multilayers, and both skyrmion modes are tunable through the multilayer geometry.  
%}
%Notably, our multilayers enable the detection of a 
%disting CCW rotation of individual skyrmions up to room temperature. Furthermore, we uncover 
%% a novel 
%\textcolor{blue}{an additional}
%collective HF mode which also possesses a CCW character. Both skyrmion modes are tunable through the multilayer geometry.  
%
We note that our numerical calculations also predict a breathing mode in the multilayer with lower frequency  $\sim$1.5\,GHz.  However this can only be probed with OP ac magnetic fields (Supplementary section II.F) and is beyond the experimental scope of this study. \\
In summary, we have studied the dynamic properties of magnetic skyrmions in technologically relevant multilayers hosting N\'eel skyrmions at room temperature. Broadband microwave absorption measurements reveal the presence of a low frequency resonance associated with CCW gyration of isolated magnetic skyrmions. The interlayer dipolar interaction in our multilayers generates a high frequency collective skyrmion resonance with CCW character. Crucially, these resonant modes can be tuned over a broadband frequency range by minor adjustments in the stack geometry. Combined with our earlier results~\cite{Soumyanarayanan2017}, we have now established that both static and dynamic properties of skyrmions can be tuned via the multilayer stack architecture. Our results expose the potential of skyrmion-hosting multilayers as a material platform capable of combining conventional dc spintronics (for data storage and logic operations) with magnonic or synaptic devices operating at GHz frequencies.

 \clearpage

\textbf{Acknowledgements}: We acknowledge N.K. Duong and A.K.C. Tan for preliminary MFM images, and A. Soumyanarayanan for input in the early stages of the project. This work was supported by Singapore National Research Foundation (NRF) Investigatorship (Ref. No.: NRF-NRFI2015-04) and the Ministry of Education (MOE), Singapore Academic Research Fund Tier 2 (Ref. No. MOE2014-T2-1-050), MOE AcRF Tier 3 Award MOE2018-T3-1-002 and Tier 1 Grant No. M4012006. M.R. thanks the Data Storage Institute, Singapore for access to sample growth facilities. V.K. thanks U. Nitzsche for technical support and acknowledges financial support from the Alexander von Humboldt Foundation and the National Academy of Sciences of Ukraine (Project No. 0116U003192). M.G. acknowledges financial support from DFG CRC 1143 (Project No. 247310070) and DFG Project No. 270344603 and 324327023. 

\textbf{Author Contribution}: C.P. conceived the research and coordinated the project. MR carried out the film deposition and  characterisation.  S.H. and C.P. designed the FMR apparatus.  B.S., S.H. and L.P. performed the FMR experiments.  V.K. performed the micromagnetic simulations with input from B.S., A.P.P., F.M., and C.P.  B.S., A.P.P., V.K. and C.P. analysed the data.  M.G. coordinated the theoretical part of the project.  A.P.P., B.S. and C.P. wrote the manuscript, with discussions and contributions from all authors.

\textbf{Additional Information}: Supplementary information is provided with this manuscript.

\textbf{Competing Interests}: Authors declare no competing interests.

\textbf{Data Availability}: Data are available from the corresponding authors upon reasonable request.

\textbf{Correspondence}: Correspondence and request for materials must be addressed to: \\ christos@ntu.edu.sg,  appetrovic@ntu.edu.sg

\clearpage

\section {Methods}

\subsection {Sample growth and initial characterisation}
All our [Ir/Fe/Co/Pt] heterostructures are grown on Si wafers, with a $\sim$1$ \,\mu$m thick thermally oxidised surface layer. The multilayers are grown by dc magnetron sputtering in an ultra-high vacuum chamber. The multilayer composition is Ta\textsubscript{3}/Pt\textsubscript{10}/[Ir\textsubscript{1}/Fe\textsubscript{0.5}/Co\textsubscript{0.5}/Pt\textsubscript{1}]\textsuperscript{20}/Pt\textsubscript{2}, where the subscripts indicate the layer thicknesses in nm and the magnetically active [Ir\textsubscript{1}/Fe\textsubscript{0.5}/Co\textsubscript{0.5}/Pt\textsubscript{1}] unit is stacked 20 times. The Ta\textsubscript{3}/Pt\textsubscript{10} seed layer is included to optimise the crystallinity of the upper layers. Deposition at low power (25\,W) and chamber pressure ($1.5\times 10^{-3} $\,Torr Ar) allows us to control the layer thicknesses with Angstrom sensitivity, resulting in systematic and reproducible magnetic properties of our heterostructures.
We determine the magnetic parameters of our multilayers using a combination of bulk magnetometry and local imaging. A Quantum Design\textsuperscript{TM} MPMS-XL SQUID magnetometer was used to measure the saturation magnetisation $M_s$.  Spatially resolved magnetic imaging was performed using a Bruker\textsuperscript{TM} D3100 atomic force microscope equipped with Nanosensors\textsuperscript{TM} ultra-low moment magnetic tips (radius $<$ 15\,nm) and a homogeneous, adjustable perpendicular field provided by a permanent magnet array. Images were typically acquired with a tip lift height of 20\,nm, which is sufficient to prevent the stray field of the tip from influencing the chiral spin textures.

\subsection{Ferromagnetic Resonance Measurements}
We acquire field-sweep and frequency-sweep microwave absorption spectra via a broadband technique, previously used to measure resonance in a range of thin film magnets. Our heterostructures are secured upside-down on a coplanar waveguide using a spring-loaded clamp. This waveguide is mounted on the cold finger of a variable-temperature probe in an electromagnet, and the microwave transmission $ S_{12}$ is measured using a Keysight\textsuperscript{TM} PNA N5222 vector network analyser (VNA) in two-port mode. The useful bandwidth of this technique (primarily limited by the length of the coaxial cables) extends above 30\,GHz, far beyond the frequency range studied here. Field-swept measurements exhibit lower noise (and a larger normalised drop in $ S_{12}$ at resonance), due to the limited response time and hence reduced sensitivity of the VNA during frequency sweeps.
The measured FMR spectra contain superposed absorptive and dispersive components, the latter contributing to the phase lag between the microwave excitation field and the magnetisation response~\cite{WIrthmann2010}. Following Dyson’s original approach to modelling spin resonances~\cite{Dyson1955} and standard practices in the FMR community~\cite{Mecking2007, Harder2011, Nembach2011} we fit all our experimental spectra with a superposition of symmetric and antisymmetric functions:
\begin{equation} 
S_{12} (x)=A \Delta x \frac{\Delta x+\beta(x-x_0 )}{(x-x_0 )^2}+(\Delta x)^2 +Dx+C     
\end{equation}           
where $A$ is the amplitude of the (Lorentzian) absorption component, $\beta$ is the dispersion to absorption ratio, $D$ accounts for a linear drift in the VNA output over time~\cite{Nembach2011}  and $C$ describes the constant background signal from cable losses. Resonance occurs at $x=x_0$ with spectral linewidth $\Delta x$, where $x \equiv f $ for frequency sweeps and $x \equiv H$ for field-swept data. 

\subsection{Micromagnetic Simulations}
We model the local magnetisation and microwave response of our multilayers using the MuMax\textsuperscript{3} software package~\cite{Vansteenkiste2014}. 
This software solves the Landau-Lifshitz equation
$\partial_t {\bf m} = - \frac{\gamma}{1+\alpha^2} ({\bf m} \times {\bf B}_{\rm eff} + \alpha {\bf m} \times ({\bf m} \times {\bf B}_{\rm eff}))$ for the unit magnetisation vector ${\bf m} = {\bf M}/M_s$ with gyromagnetic ratio $\gamma$, damping constant $\alpha$ and saturation magnetisation $M_s$. The random stochastic field which can be used to simulate thermal fluctuations was switched off in order to avoid unnecessary noise. 
The effective magnetic field ${\bf B}_{\rm eff} = - \delta E/\delta {\bf M}$ is derived from the magnetic energy $E$, which is explicitly shown in the supplementary materials (section III.A) and includes the exchange interaction $A$, DMI $D$, uniaxial anisotropy $K$, Zeeman energy and dipolar interaction.
The parameters $A$ =  9.25 pJ/m, $D$ = 1.40 mJ/m\textsuperscript{2}, $ K$ = 0.65 MJ/m\textsuperscript{3} and $M_s$ = 1.02 MA/m  were adjusted to provide the best fit to our combined experimental magnetometry, MFM and FMR data (see  Supplementary section I.A). The discretisation is chosen in the form of a cuboid with dimensions $\Delta x\times\Delta y\times\Delta z$, where $\Delta z$ is set to 1\,nm for all geometries (except for the thickness dependence study in Fig.~\ref{fig:5} where $\Delta z$ = 0.5\,nm for $L$ = 0.5\,nm). For the large 2$\times$2 $\mu$m\textsuperscript{2}  geometry simulations of Fig.~\ref{fig:3} the in-plane discretisation is $\Delta x=\Delta y=3.9$ nm, which is reduced to 1\,nm for the additional simulations in Fig.~\ref{fig:4} and~\ref{fig:5}. We simulated a ferromagnetic-nonmagnetic [FM (1\,nm) - NM ($L$)] bilayer with $L$ = 2\,nm corresponding to the experimental setup. The thickness $L$ was also varied  for the simulations presented in Fig.~\ref{fig:5}e,f,h,i. Periodic boundary conditions (PBC) were applied in all three directions, thus approximating the 20 bilayers of the experimental geometry. For every magnetic field step in our simulations, the system was relaxed from an initial random magnetisation state. This yields disordered skyrmion lattice configurations which are metastable, since they can be annealed into an ordered  skyrmion lattice by the repeated application of oscillating magnetic field pulses. As these metastable configurations resemble our experimental observations, they have been used to obtain the net magnetic moment, the magnetisation maps and the microwave absorption spectra shown in Fig.~\ref{fig:3}. For the spectra, we applied a sinc pulse of an excitation magnetic field $\textbf{b}(t) $= ${\hat{\textbf{e}} _x} b_0$sinc$(2\pi f_{max} (t-t_0))$ with  $f_{max}$ = 20\,GHz, $t_0$=1\,ns and $b_0$ = 10\,mT, then Fourier transformed the induced in-plane magnetisation oscillations $m_x$. For the sake of comparability with the experimental data, we show the absolute value of the Fourier transform. Noise at small frequencies is tentatively attributed to the relaxation processes of the metastable state. The simulations in Fig.~\ref{fig:4} were performed on an in-plane 1$\times$1 $\mu$m\textsuperscript{2} area with a metastable configuration comprising a single skyrmion in a field-polarised background; a sinc pulse with $b_0$ = 2\,mT was applied to generate the absorption spectra. Finally, the simulations of Fig.~\ref{fig:5} were performed on an in-plane area corresponding to a unit cell of a hexagonal skyrmion lattice, where the lattice constant was adjusted as described in the main text. Here, a sinc pulse with $b_0$ = 1\,mT was applied to obtain the absorption spectra. 

\end{document}